\documentclass{pasj00}

\SetRunningHead{A. Arai et al.}{Optical and Near-Infrared Photometry of V2362~Cyg}
\Received{2009/03/14}
\Accepted{2010/05/24}
\Published{}
\begin{document}

\title{
Optical and Near-Infrared Photometry of Nova V2362~Cyg :
Rebrightening Event and Dust Formation
}

\author{
Akira \textsc{Arai}\altaffilmark{1,
\thanks{Present address: Koyama Astronomical Observatory, Kyoto Sangyo University,
Motoyama, Kamigamo, Kita-ku, Kyoto, 603-8555}}
Makoto \textsc{Uemura}\altaffilmark{2},
Koji S. \textsc{Kawabata}\altaffilmark{2},
Hiroyuki \textsc{Maehara}\altaffilmark{3},
Kazuhiro \textsc{Nakajima}\altaffilmark{4},\\
Seiichiro \textsc{Kiyota}\altaffilmark{5},
Taichi \textsc{Kato}\altaffilmark{6},
Takashi \textsc{Ohsugi}\altaffilmark{1,2},
Takuya \textsc{Yamashita}\altaffilmark{2,   
\thanks{Present address: National Astronomical Observatory, 2-21-1
Osawa, Mitaka, Tokyo 181-8588}},
Mizuki \textsc{Isogai}\altaffilmark{1,$^*$},
Osamu \textsc{Nagae}\altaffilmark{1},\\
Shingo \textsc{Chiyonobu}\altaffilmark{1},
Yasushi \textsc{Fukazawa}\altaffilmark{1},
Tsunefumi \textsc{Mizuno}\altaffilmark{1},
Hideaki \textsc{Katagiri}\altaffilmark{1}, 
Hiromitsu \textsc{Takahashi}\altaffilmark{2},\\
Kiichi \textsc{Okita}\altaffilmark{8}, 
Michitoshi \textsc{Yoshida}\altaffilmark{8},
Kenshi \textsc{Yanagisawa}\altaffilmark{8}, 
Shuji \textsc{Sato}\altaffilmark{9},
Masaru \textsc{Kino}\altaffilmark{9},\\
Masahiro \textsc{Kitagawa}\altaffilmark{9} and
Kozo \textsc{Sadakane}\altaffilmark{10}
}
\altaffiltext{1}{
Department of Physical Science, Hiroshima University, 
Kagamiyama 1-3-1, Higashi-Hiroshima 739-8526}
\email{arai6a@cc.kyoto-su.ac.jp}
\altaffiltext{2}{
Hiroshima Astrophysical Science Center, Hiroshima University, Kagamiyama
1-3-1, Higashi-Hiroshima 739-8526}
\altaffiltext{3}{
Kuwasan Observatory, Kyoto University
17 Ohmine-cho Kita Kazan, Yamashina-ku, Kyoto City, Kyoto, 607-8471
}
\altaffiltext{4}{VSOLJ, 124 Isatotyo Teradani, Kumano, Mie 519-4673}
\altaffiltext{5}{VSOLJ, Azuma 1-401-810, Tsukuba 305-0031}
\altaffiltext{6}{
Department of Astronomy, Kyoto University, Sakyo-ku, Kyoto 606-8502}
\altaffiltext{7}{
Okayama Astrophysical Observatory, National
Astronomical Observatory of Japan, Kamogata Okayama 719-0232}
\altaffiltext{8}{
Department of Physics, Nagoya University, 
Furo-cho, Chikusa-ku, Nagoya 464-8602}
\altaffiltext{9}{
Astronomical Institute, Osaka Kyoiku University,
Asahigaoka, Kashiwara, Osaka 582-8582}


\maketitle

\begin{abstract} 
We present optical and near-infrared (NIR) photometry of a classical
nova, V2362 Cyg (= Nova Cygni 2006). V2362 Cyg experienced a peculiar
rebrightening with a long duration from $100$ to $240$ d after the
maximum of the nova.  Our multicolor observation indicates an emergence
of a pseudophotosphere with an effective temperature of $\sim 9000$~K
at the rebrightening maximum.  After the rebrightening maximum, the
object showed a slow fading homogeneously in all of the used bands for one
week. This implies that the fading just after the rebrightening maximum
($\lesssim 1$ week) was caused by a slowly shrinking pseudophotosphere.
Then, the NIR flux drastically increased, while the optical flux
steeply declined.  The optical and NIR flux was consistent with
blackbody radiation with a temperature of $\sim1500\;{\rm K}$ during
this NIR rising phase.  These facts are likely to be explained by dust
formation in the nova ejecta.  
Assuming an optically thin case, we estimate the dust mass of 
$10^{-8}-10^{-10}$ M$_{\solar}$, which is less than those in typical
dust-forming novae.
These results support the senario that a second, long-lasting outflow, 
which caused the rebrightening, interacted with a fraction of the initial 
outflow and formed dust grains.
\end{abstract}

\section{Introduction} 

Classical novae are cataclysmic variable stars which are semidetached 
binary systems containing a late-type star and a white dwarf (WD). 
The nova outburst is induced by a thermonuclear runaway in the matter
transferred from the late-type star onto the surface of the WD. 
A large amount of gas accumulated on the WD is ejected due to nova
outbursts. The ejected mass is typically $\sim 10^{-4}$ M$_{\odot}$ in
an outburst. The expansion velocity of the ejecta reaches 
$\sim 10^{2}$--$10^{3}{\rm km}\;{\rm s}^{-1}$. 
The outburst amplitude of novae is $10$--$15$ mag in optical
\citep{bode_evans89}. 

Dust formation episodes are occasionally observed during nova outburst phases.
The dust formation occurs 30 -- 80 d after the maximum of the light
curve. This phase is the ``transition phase'' \citep{bode_evans89}.  
We can study the formation and cooling processes of dust through the 
temporal evolution of novae.

V2362~Cyg (= Nova~Cygni~2006) was discovered on April 2.807 UT by
H. Nishimura at $10.5\,{\rm mag}$ \citep{nakano06_iauc8697}. 
\citet{yamaoka06_iauc8698} spectroscopically confirmed that it was a
classical nova just around its maximum from the H$\alpha$ emission line
with a clear P-Cygni profile.  \citet{siviero06_iauc8702} reported that
the nova belongs to the Fe\emissiontype{II} class.  \citet{steeghs06}
identified the progenitor of V2362~Cyg on their galactic plane survey 
images taken on 2004 Aug 3 as a point source with the
magnitudes of $r'=20.3 \pm 0.05$ and $i'=19.76 \pm 0.07$, which
indicates the outburst amplitude of $\sim 12\,{\rm mag}$.

The nova experienced an unusual rebrightening event from $\sim 100$ d
after the first maximum. 
The nova reached its rebrightening maximum on 2006 December 1.9 (= JD
2454071.4; \cite{munari08}). 
A similar rebrightening event has been found only in V1493~Aql 
\citep{bonifacio00,venturini04}. V2362~Cyg is the second case which exhibited such a 
rebrightening, and hence, has received much attention  
\citep{goranskij06,munari06_cbet671,lynch06_iauc8785,rayner06_iauc8788,kimeswenger08,lynch08,munari08,
poggiani09}. 
Most recently, V2491~Cyg showed a small rebrightening event and some 
authors suggested that V2491~Cyg belongs to the same class 
\citep{naik09,hachisu09,takei09}. Such rebrightening events are totally 
unexpected from the standard picture of classical novae, and their 
mechanism is still unclear. Optical--near-infrared multiband monitoring would be 
crucial to explore the temporal evolution of the photosphere and outflow 
material around the rebrightening event.

We performed extensive optical and near-infrared (NIR) photometric 
observations of V2362~Cyg covering the early decline and the 
rebrightening event of the object.  Our data trace the evolution of 
outflow materials around the rebrightening event.  In section~2, we describe 
our observations.  The results are presented in section~3. We discuss the 
characteristics of the dust formation in section~4. Finally, we summarize our 
findings in section~5. 

\section{Observations and Data Reduction}

We performed simultaneous optical and NIR observations during the
rebrightening event with TRISPEC attached to the KANATA 1.5-m telescope
at Higashi-Hiroshima Observatory. TRISPEC is an imager and spectrograph
with a capability of polarimetry covering both optical and NIR
wavelengths (\cite{watanabe05}).
The TRISPEC observation was conducted from 2006 November 20.56 UT
(JD~2454060.06), 11 d before the rebrightening maximum, to 2007
February 5.87 (JD~2454137.37), 68 d after past the rebrightening maximum. 
We obtained typically 10 frames in a night with different dithering
positions in individual bands. 
Exposure times in the $B$, $V$, $R_{\rm c}$, $I_{\rm c}$, $J$, $H$ and $K_{\rm s}$
in a night were typically 450, 380, 150, 155, 140, 180 
and 75 s, respectively. We have no $J$- and $H$-band images between 2006
December 15 and 2007 January 18 because of a hardware problem.

We obtained differential magnitudes of the object relative to a
comparison star using aperture photometry with the APPHOT package in
IRAF{\footnote {IRAF is distributed by the National Optical Astronomy
Observatory,  which is operated by the Association of Universities for
Research in Astronomy, Inc., under cooperative agreement with the
National Science Foundation.}}.  
We used a nearby star TYC~3181-1159-1 as a comparison star. 
Its magnitudes of $B=11.15\pm0.01$ and $V=9.70\pm0.01$ are quoted 
from \citet{frigo06} and $J=7.18\pm0.02$, $H=6.50\pm0.02$ and 
$K_{\rm s}=6.35\pm0.02$ are from the 2MASS catalog 
\citep{skrutskie06}. We estimated the magnitude of the comparison star
at $R_{\rm c}=8.95\pm0.02$ and $I_{\rm c}=8.26\pm0.03$, using the other
nearby stars listed in \citet{frigo06}.

In addition to the TRISPEC/KANATA observation, we performed optical CCD
photometry of V2362~Cyg from 2006 April 5.8 (JD~2453831.3), around the
fist maximum, to 2007 January 23.4 (JD~2454123.9) with several telescopes in 
30-cm class. For differential photometry, we used the nearby stars,
TYC 3181-1511-1 and TYC 3181-1401-1 taken in the same frames including 
the nova.

For the correction of interstellar extinction, we used $E_{B-V} = 0.58
\pm0.02$, which is an average of the values reported in
\citet{siviero06_iauc8702}, \citet{russell06_iauc8710} and
\citet{mazuk06_iauc8731}.  We calculated the interstellar extinction in
each band using the reddening curve presented in
\citet{rieke_lebofsky85} and \citet{schlegel98}.

\section{Results}
\begin{figure}[h!]
  \begin{center}
    \FigureFile(85mm,80mm){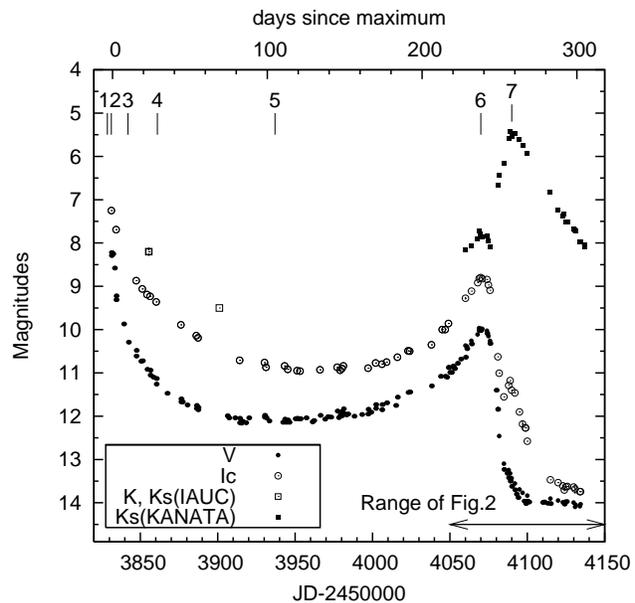}
  \end{center}
  \caption{Optical and NIR light curves of V2362~Cyg from 2006 April to 2007
 February. The NIR data on 2006 April 30 (JD~2453855.5) and 2006 June 14.6
 (JD~2453901.1) are quoted from \citet{russell06_iauc8710} and 
\citet{mazuk06_iauc8731}, respectively. The lower horizontal axis indicates
 the time in JD. In the upper axis, we show the days
 from the first maximum, 2006 Apr 5.8 UT (JD~2453831.3). The vertical
 ticks indicate notable epochs shown in Table~\ref{LCnote}.
}\label{lc}
\end{figure}

\begin{figure}[h!]
  \begin{center}
    \FigureFile(80mm,60mm){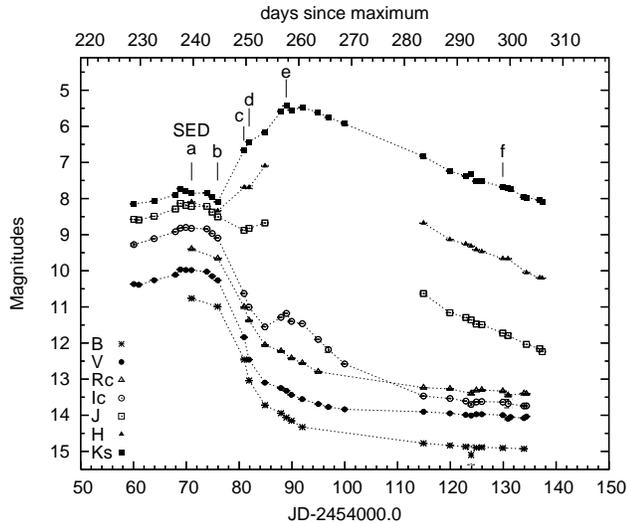}
  \end{center}
  \caption{Light curves before and after the rebrightening maximum.
    The six vertical lines indicate epochs of spectral energy
    distribution (SED) shown in figure~\ref{sed}.
  }\label{kanata}
\end{figure}

Figure~\ref{lc} and \ref{kanata} show the light curves of V2362~Cyg.
The latter exhibits the light curves after the rebrightening maximum in detail.

\subsection{The First Maximum and Early Decline}

The nova reached its first maximum at V=8.2 on 2006 April 5.8 in our data.  
In this paper, we denote the fiducial time by $t$, in days, where $t=0$ is 
defined at the first maximum.  We estimated $t_2=10.5 \pm 0.5$ 
and $t_3=29 \pm 1$ from our $V$-band light curve, where $t_2$ and 
$t_3$ are the time taken for the nova to fade by 2 and 3~mag from 
the maximum, respectively. 

Our $t_3$ is slightly longer than those in other studies
(\cite{kimeswenger08}; \cite{munari08}; \cite{poggiani09}). This is
because our maximum magnitude $V=8.2$ is fainter than others ($V=7.8$--$8.0$) 
due to the sparseness of our observation around the maximum.  Even in $t_3$ 
estimated with our data, V2362~Cyg is included in the fast nova, as reported 
in the other studies. 

In the early decline phase ($t=0$--$100$), the nova showed a  monotonous decline 
without a transition phase, as can be seen in figure~\ref{lc}.  This figure also 
shows the NIR observations reported in \citet{russell06_iauc8710} and \citet{mazuk06_iauc8731}.   
We estimated the reddening-corrected color $V-K_{\rm s}=1.2 \pm 0.1$ and 
$V-K=0.8 \pm 0.1$ on $t=24$ and $t=70$, respectively.  
The characteristics of the light curve and color were well consistent with 
those of typical fast novae.   

\begin{table}[!b]
\begin{center}
\caption{Characteristic epochs of V2362~Cyg.}
\label{LCnote}
\begin{tabular}{cllc}              
\hline
{\small Epoch} & {\small Remark}  &  {\small $t$ days} & {\small $V$-mag}\\ 
 \hline
{\small $1$} & {\small Discovery}  &  {\small$-3.0$\footnotemark[$*$]} & {\small 10.5\footnotemark[$*$]}\\ 
{\small $2$} & {\small First maximum}                & {\small$0$} & {\small 8.2}\\
{\small $3$} & {\small $t_{2}$\footnotemark[$**$]}   & {\small$10.5_{\pm 0.5}$}\\
{\small $4$} & {\small $t_{3}$\footnotemark[$**$]}   & {\small$29_{\pm 1}$} \\
{\small $5$} & {\small Onset of rebrightening}       & {\small$\sim 100$} & {\small 12.1}\\ 
{\small $6$} & {\small Rebrightening maximum}        & {\small$240$\footnotemark[$\dagger$]} & {\small 10.0}\\  
{\small $7$} & {\small $K_{\rm s}$ maximum}     & {\small$258$} & {\small 13.3}\\
 \hline
 \multicolumn{4}{@{}l@{}}{\hbox to 0pt{\parbox{80mm}{\footnotesize
  {\small \footnotemark[$*$] Using photographic films \citep{nakano06_iauc8697}.  
 \footnotemark[$**$] $t_{n}$ is at the date which the nova faded by
 $n$ magnitudes from the maximum in $V$-band.
 \footnotemark[$\dagger$] The date of the rebrightening maximum is JD 2454071.4.
}\hss}}}
\end{tabular}
\end{center}
\end{table}

\subsection{Rebrightening and Subsequent Decline}

The unusual rebrightening started on $t\sim 100$, as
can be seen in figure~\ref{lc}.  The slope of the light curve gradually
increased until $t=240$ when the object reached the apparent maximum at
$V=10.0$.   

\begin{figure}[t]
  \begin{center}
    \FigureFile(80mm,50mm){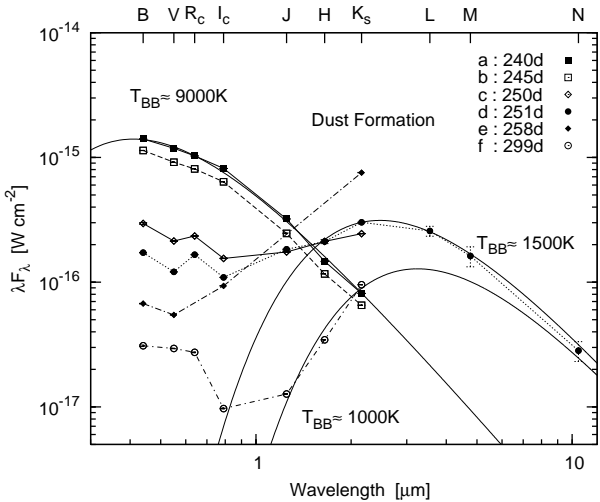}
  \end{center}
  \caption{Optical and NIR SED of V2362 Cyg at
    six epochs. The solid lines indicate single-temperature blackbody radiations
    ($9000\;{\rm K}$, $1500\; {\rm K}$ and $1000\; {\rm K}$). The $L$-, $M$-
    and $N$-band data on $t=251$ (epoch d) were quoted from
    \citet{lynch06_iauc8785}. Epoch `a' is almost at the rebrightening
    maximum and `e' is at the $K_{\rm s}$-band maximum. The interstellar
    extinction was corrected as described in \S~2.} 
  \label{sed}
\end{figure}

Our multi-band photometric observations allow us to 
study the spectral energy distribution (SED) of the optical--NIR 
regime during the rebrightening.  Figure~\ref{sed} shows the 
observed SEDs.  In this figure, we draw 6 SEDs at epoch (a, b, c, d, 
e, and f) which are shown in figure~\ref{sed} and 
also indicated in figure~\ref{kanata}.  At epoch~a, the observed 
SED can be described with a blackbody emission as shown in figure~3. 
We derived the best-fit temperature at $T=9000 \pm 200\;{\rm K}$.  
The SED, hence, indicates that the optical--NIR flux was dominated 
by an optically thick pseudophotosphere at the rebrightening maximum.

In ordinary novae, the blackbody-like photospheric emission dominates 
the optical flux only around the maximum light.
In the case of V2362~Cyg, \citet{czart06} reported the optical spectrum 
during the earliest phase, and suggested that its blue spectrum resembled 
that of an A5--A7 supergiant.  This indicates that the optical emission was 
dominated by a pseudophotosphere with $T=8000$--$8600\;{\rm K}$ at the 
outburst maximum \citep{cox00}.  
The photospheric temperature at the rebrightening maximum was,
therefore, comparable to or slightly higher than, that at the 
first maximum, while the rebrightening maximum was fainter.
Assuming blackbody emission, we can readily compare the sizes of 
the emitting region at the initial and rebrightening maxima using  
the apparent $V$-magnitude and the temperatures.  As a result, 
the diameter of the photosphere at the rebrightening maximum is 
estimated to be about half of that at the first maximum.

\begin{figure}
  \begin{center}
    \FigureFile(80mm,60mm){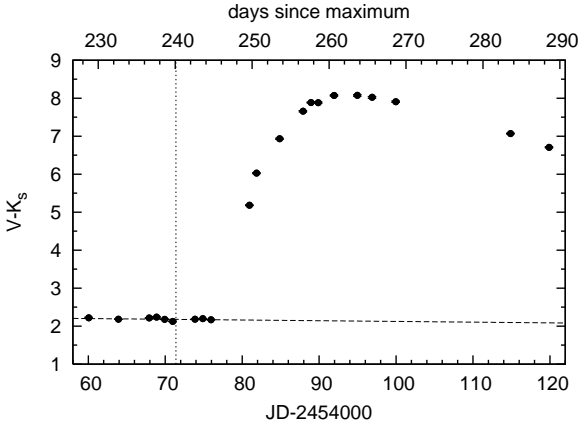}
  \end{center}
  \caption{$V-K_{\rm s}$ color before and after the rebrightening
    maximum. The interstellar reddening is not corrected for.
    The vertical dotted line indicates epochs of the rebrightening
    maximum. The horizontal dashed line indicates the best-fitted line of
    $V-K_{\rm s}$ between $t=229$ and $t=245$. 
  }\label{color}
\end{figure}

After the rebrightening maximum, the object started a gradual fading in
all optical--NIR bands. The $V-K_{\rm s}$ color just before and after
the rebrightening maximum is shown in figure~\ref{color}. There was no
significant  change across the rebrightening maximum, i.e., in $t=$228--245. 
This suggests that the shape of the SED remained almost constant irrespective
of the flux variation across the rebrightening maximum ($\Delta V\sim 0.5$).

Sudden brightenings in $H$- and $K_{\rm s}$-bands were then detected 
on $t=250$ (epoch c).  
As shown in figure~\ref{sed}, the pseudophotosphere component rapidly
decreased, while another NIR component increased around the epoch.  The
brightening was observed in $J$- and $I_c$-bands a few days
later.  In figure 3, we plot $L$-, $M$-, and $N$-band data on 
$t=251$ (epoch d) reported by \citet{lynch06_iauc8785} and also give
blackbody spectra with temperatures of 1500 and 1000~K.  
The 1500~K blackbody spectrum well explains the observed IR SED on $t=251$. 
Hence, this IR emission can be interpreted as the thermal
emission from dust grains.  \citet{munari08} and \citet{lynch08}
reported the dust temperature of $1550\;{\rm K}$ on $t=251$ and
$1410\;{\rm K}$ on $t=260$, respectively. The dust shell would be
formed, presumably associated with the rebrightening event. The rapid
decrease of the pseudophotosphere component is likely to be connected
with the increasing absorption by the dust shell.

After the  $K_{\rm s}$-band maximum on $t=258$ (epoch e), 
the object faded in all optical--NIR bands.  At $t=299$ (epoch f), 
as shown in figure~\ref{sed}, the NIR emission can be described 
with the 1000~K blackbody spectrum.  \citet{lynch07_iauc8849} reported
the dust temperature was below $ < 520$~K on $t \sim 422$.  
The dust cooling, thus, seems to have started just after the 
$K_{\rm s}$-band maximum, probably due to cessation of the
dust formation and expansion of the dust shell.

\section{Discussion}

\subsection{Estimation of the Dust Mass}

Assuming an optically thin dust cloud and a distance to V2362~Cyg of
$d=7.2$~kpc, \citet{munari08} estimated the mass of carbonaceous dust
grains of $M \sim 10^{-9}$M$_{\solar}$.
However, they underestimated the dust mass because they used the NIR flux 
at pre-maximum phase (epoch d; 7 days before the $K_{\rm s}$ maximum)  
when the bolometric luminosity of the dust cloud  ($L_{\rm dust}$) was 
still 0.4 times as large as the maximum value on epoch e.
The uncertainty of the distance should also be considered for the error 
estimation of the dust mass.
Then, we estimate the dust mass of V2362~Cyg with our data under the same 
assumption as \citet{munari08} that carbon dusts were formed after the rebrightening. 
Although the late time observation ($t=443$) suggests that the IR
spectrum was not typical for carbonaceous dust emission \citep{lynch08},
the dominant species could be changed with epochs 
(e.g., QV~Vul, \cite{gehrz92}; V705~Cas, \cite{mason98})

According to \citet{woodward93}, 
the mass of the dust cloud ($M_{\rm dust}$) is derived as
{\footnotesize
\begin{eqnarray}
M_{\rm {dust}} & = & \frac{4\pi}{3} N \rho a^3 \nonumber \\
& \simeq & 1.1 \times 10^6 \left( \frac{d}{1\mbox{ kpc}} \right)^{2}
\left[ \frac{(\lambda F_{\lambda})_{\rm max}}{1\mbox{ W cm}^{-2}}\right]
 \left( \frac{T_{\rm dust}}{1000\mbox{ K}} \right)^{-6} \mbox{M}_{\solar},
\end{eqnarray}
}
where $d$ is the distance to the nova, N is a number of total dust grains and $T_{\rm dust}$ 
is a temperature of dust grains. We assume graphite dust grains of a density of $\rho = 2.25$ 
and a Planck-mean cross section coefficient $Q \sim 10^{-2} aT^2$ 
(for smaller grains with size of $a \lesssim 0.5$ $\micron$; \cite{gilman74,woodward93}). 
Adopting $T_{\rm dust}=1500$ K estimated from the IR data in $t=251$ (epoch~d) and 
$(\lambda F_{\lambda})_{\rm {max}} = 7.6 \times 10^{-16}$~Wcm$^{-2}$ 
in our data at NIR maximum ($t=258$, epoch e), we obtain
{\footnotesize
\begin{equation}
M_{\rm {dust}} \sim 7.3\times 10^{-11} 
\left( \frac{d}{1\mbox{ kpc}} \right)^2 \mbox{ M}_{\solar}.
\end{equation}
}
We note that the deriving mass does not mean the whole mass formed in the 
ejecta after the rebrightening event, but the mass evaluated at the NIR 
maximum ($t=258$, epoch~e).

$M_{\rm dust}$ highly depends on the distance $d$,
although it has not been determined well; $d=1.5$~kpc \citep{czart06},
$5$--$12$~kpc \citep{steeghs06}, $5.5$--$10.0$~kpc \citep{kimeswenger08},
$7.2 \pm 0.2$~kpc \citep{munari08} and $7.2$--$15.8$~kpc
\citep{poggiani09}.
For the nearest ($d=1.5$ kpc) and the most distant ($d=15.8$~kpc)
cases, the $M_{\rm dust}$ becomes $\sim 2\times 10^{-10}$ M$_{\odot}$
and $2\times 10^{-8}$ M$_{\odot}$, respectively.
Even if we take the uppermost value, $M_{\rm dust}$ is comparable to or
smaller than those of ordinary dust-forming novae
($10^{-6}$--$10^{-8}$~M$_{\solar}$; e.g. \cite{gehrz98_review}).

It is noted that the dust mass should be larger if the dust cloud
is optically thick as indicated by \citet{lynch08}. 
If this is the case, the derived $M_{\rm dust}$ should be the
lower-limit and the actual dust mass would be larger. On the other hand,
our NIR data suggests that $L_{\rm dust}$ is only $0.4$--$0.5$ times
as large as the bolometric luminosity of nova ($L_{\rm nova}$)
even at the $K_{\rm s}$ maximum, if we assume 
$L_{\rm nova}=1.3\times 10^{38}$ erg s$^{-1}$ 
with $d=7.2$~kpc as \citet{munari08}. 
$L_{\rm dust}$ is calculated by integrations of blackbody SEDs fitted
with $K_{\rm s}$-band data assuming dust temperatures with a rage of
$T=1410$--$1500$~K and $d$=7.2~kpc.
These facts suggest that the dust cloud does not fully cover the nova
(i.e., covers by less than 4$\pi$ str), if the dust cloud was optically
thick. A patchy cloud is preferable. 

\subsection{Rebrightening and Dust Formation in the Second Mass-Flow}

\citet{munari06_cbet739} and \citet{kimeswenger08} detected P-Cygni
profiles in the optical spectrum of V2362~Cyg during its rising phase
to the rebrightening maximum ($t\sim 170$).  In conjunction with the
existence of the blackbody emission of $T \sim 9000$ K, the rebrightening
phase is reminiscent of the maximum phase of ordinary novae
(\cite{bode_evans89}).
We propose that the rebrightening event was caused by the reformed
pseudophotosphere by a delayed mass-outflow. \citet{hachisu09} recently
proposed that the energy source of the rebrightening is magnetic
reconnections around the pseudophotosphere formed by the first outflow.
Such a magnetic activity may have accelerated the gas and caused the
second outflow in V2362~Cyg. 

As discussed in section~4.1, the mass of the dust grains formed after the
rebrightening is likely to be smaller than those in other dust-forming
novae. If the dust grains are formed in the first ejecta cooled predominantly
by expansion (i.e., classical dust formation mechanism, e.g.,
\cite{clayton76}) the epoch of the dust formation is expected to be
earlier ($t\simeq 30-100$ d) and the dust mass should be larger, as
in ordinary dust-forming novae. 

The ejecta of the second outflow would contain less materials than that of
the initial outflow because the peak luminosity of the rebrightening was
much lower than the initial peak. 
Thus, masses are considered to be small for both the dust and the ejecta 
in the second outflow. 
This suggests that the dust grains would be formed preferentially 
from the material of the second outflow, if the \citet{lynch08}'s 
scenario that the dust formation would be triggered by the interaction 
between the second outflow and the initial flow is the case.

\subsection{The Termination of the Rebrightening}

Here we briefly discuss the nature of the termination of the
rebrightening, i.e., the beginning of the decline on $t\sim 240$.
We can consider two scenarios; 
(i) the rebrightening was terminated due to the increasing absorption for
the optical--NIR flux by the dust grains, and 
(ii) the rebrightening was terminated by a decay of the emission from the
pseudophotosphere itself. If the former scenario is ture, we could expect the 
reddening of the object just after the rebrightening maximum. As shown
in figure~\ref{color}, however, no reddening was detected for the first
4~d just after the rebrightening maximum. The observation, hence,
favors the latter scenario. 
If scenario (ii) is the case, the dust grains are likely to begin to
be formed about a week after the rebrightening maximum.
This implies that the dust formation might be promoted by the decay of
the emission from the shrinking pseudophotosphere.
The optical steep decline with the NIR brightening after 
$t \sim 250$ suggests that the increasing absorption due to dust formation
is significant.  However, the shrinking pseudophotosphere might also
have partly caused the decline of the optical flux (and a part of NIR one).
A full analysis with a radiation transfer for the pseudophotosphere and
the dusty ejecta would be required to reproduce the observed light curves,
which is out of the scope of this paper.

\section{Summary}

We carried out optical and NIR photometry of Nova V2362~Cyg for $\sim
300$ days from the first maximum.  Our observation indicates the
re-appearance of pseudophotosphere with $T \sim 9000$~K around the
rebrightening maximum.  The NIR flux began to increase a week after the
rebrightening maximum, accompanied with the steep decline of the optical
flux. The $K_{\rm s}$-band flux peaked at $t=258$. 
The NIR flux at the peak is well explained by the thermal dust emission with
$T_{dust} \sim 1500$~K. The dust temperature then decreased from $1500$~K to
$\sim 1000$~K within 40 days.  We estimate the dust mass of 
$M_{\rm dust} \sim 2\times 10^{-10}$--$2 \times 10^{-8}$~M$_{\odot}$.  
This support that the dust would be formed preferentially from the material 
of the second outflow, if the dust formation would be triggered by the interaction 
between the second flow and initial one as suggested in \citet{lynch08}.
\\
\\
This work was partly supported by a Grand-in-Aid from the Ministry 
of Education, Culture, Sports, Science, and Technology of Japan 
(19740104, 17340054).

\end{document}